\documentclass{article}

\usepackage{arxiv}

\usepackage[utf8]{inputenc} 
\usepackage[T1]{fontenc}    
\usepackage{hyperref}       
\usepackage{url}            
\usepackage{booktabs}       
\usepackage{amsfonts}       
\usepackage{nicefrac}       
\usepackage{microtype}      
\usepackage{amsmath,amssymb}
\usepackage{cleveref}       
\usepackage{natbib}
\usepackage{doi}
\usepackage[pdftex]{graphicx}
\usepackage[subrefformat=parens]{subcaption}
\usepackage{url}

\title{A Fast Control Plane for a Large-Scale and High-Speed Optical Circuit Switch System
\thanks{This paper extends our preliminary work published at OCP Future Technologies Symposium 
poster~\citet{Takano2021.OCP}. Specifically, we conducted additional experiments with FPGA-based
multi-master functionality.}}

\date{}

\usepackage{authblk}

\author[1]{Ryousei Takano\thanks{\texttt{takano-ryousei@aist.go.jp}}}
\author[1]{Kiyo Ishii}
\author[1]{Toshiyuki Shimizu}
\author[1]{Fumihiro Okazaki}
\author[1]{Shu Namiki}
\author[1]{Ken-ichi Sato}
\affil[1]{National Institute of Advanced Industrial Science and Technology (AIST), Japan.}

\renewcommand{\headeright}{}
\renewcommand{\undertitle}{}
\renewcommand{\shorttitle}{A Fast Control Plane for a Large-Scale and High-Speed Optical Circuit Switch System}

\hypersetup{
pdftitle={A Fast Control Plane for a Large-Scale and High-Speed Optical Circuit Switch System},
pdfsubject={cs.NI},
pdfauthor={Ryousei Takano, et. al.},
pdfkeywords={Data center network, Fast control plane, Optical Circuit Switch, Software defined network},
}

\begin{document}
\maketitle

\begin{abstract}
  We experimentally verify a fast control plane with 100 {\textmu}s of 
  configuration time that can support more than 1000 racks,
  leveraged by a software-defined network controller and 
  an industrial real-time Ethernet standard EtherCAT.
\end{abstract}


\newcommand{\ctext}[1]{\raise0.2ex\hbox{\textcircled{\scriptsize{#1}}}}


\section{Introduction}

Optical circuit switches (OCSes) are becoming increasingly important for data center networks.
Existing research works~\citet{Farrington2011.SIGCOM, Poutievski2022.SIGCOMM} employ 
OCS to enable dynamic path reconfiguration for expanding network bandwidth
and reducing power consumption.
%
K. Sato has proposed a large-scale fast OCS architecture~\citet{Sato2019.JLT} 
that efficiently combines space and wavelength routing switches.
This architecture is scalable, and it can support more than 1000 racks.
The switching time of the proposed OCS architecture is reported to be a few microseconds.
Development of a fast control plane is critical to make the best of data center 
networks that consists of several OCSes and more than 1000 electrical top of rack switches.

We have designed a fast control plane for a large-scale and high-speed OCS system.
Our previous work~\citet{Takano2021.OCP} introduces the EtherCAT technology to a control 
plane based on a Software-Defined Network~(SDN) controller.
EtherCAT is an Ethernet-based real-time fieldbus system, standardized in the IEC61158. 
EtherCAT adopts a single master and many slave architecture. 
A master periodically sends a telegram among slaves in a bucket-brigade manner,
and the slave reads and writes data on the fly.
It achieves short cycle time ($\leqq$ 100 {\textmu}s) and 
low jitter for accurate synchronization ($\leqq$ 1 {\textmu}s) while keeping
the hardware cost low.

The challenge is to achieve both high scalability and a short configuration time.
We introduce FPGA-based master and run multiple masters in parallel depending 
on the scale requirements.
This paper demonstrates the feasibility of FPGA-based multi-master functionality.
Experimental results show that low jitter with approximately 100 microseconds periodic operation  
can be achieved leveraging industry-standard and low-cost hardware and open-source software.
To the best of our knowledge, this is the first attempt to use EtherCAT to control OCSes
at data center scale.

\section{MEOW: Multi-master EtherCAT-based control plane for Optical netWorks}

\subsection{Control Plane Design}
\begin{figure}[t!]
  \centering
  \includegraphics[width=.6\columnwidth]{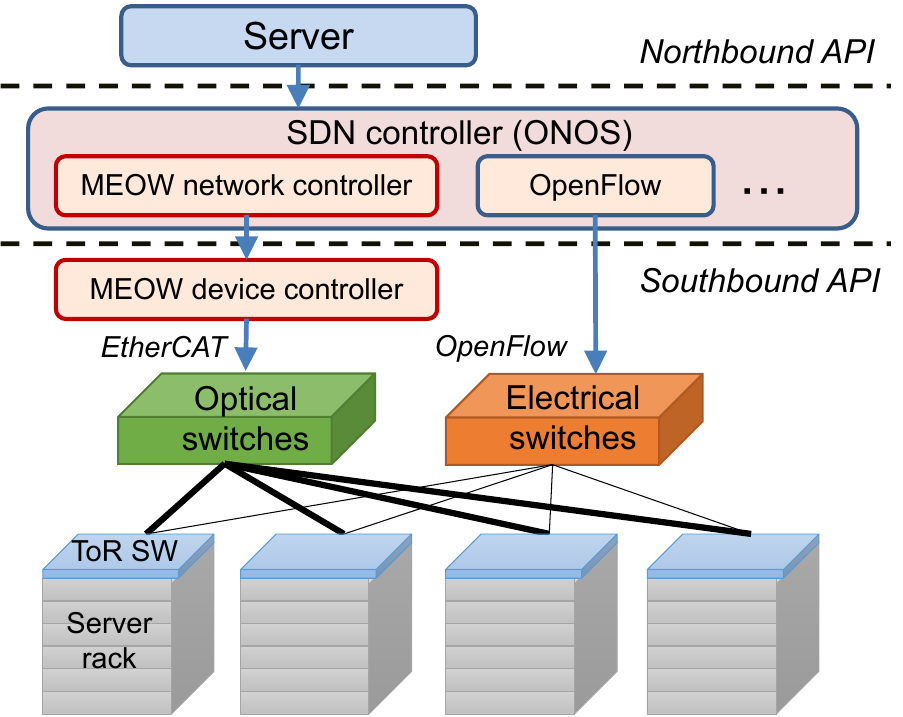}
  \caption{The overview of MEOW control plane}
  \label{fig:meow}
\end{figure}

We have designed a fast control plane based on ONOS~\citet{Berde2014.HotSDN}, 
a production-grade SDN controller platform.
ONOS is designed around extensible, modular, and distributed structure.
Figure~\ref{fig:meow} shows the schematic design of our control plane, 
a Multi-master EtherCAT-based control plane for Optical netWorks~(MEOW).

We primarily design MEOW for an optical and electrical hybrid switch network.
MEOW employs two software-driven traffic engineering approaches: service-oriented 
and data-oriented.
The former is a proactive approach and a large flow is detected according to
predefined rules by the applications.
The latter is a reactive approach and a large flow is detected based on real-time 
traffic monitoring.
In both cases, the large flow is allocated to an OCS network.
The flow switching point is placed on the top of rack switches.
OpenFlow is used for the control of electrical top of rack switches.
Additionally, we extend an EtherCAT-based network controller on top of ONOS 
for configuring OCSes.

\begin{figure}[t!]
  \centering
  \includegraphics[width=\columnwidth]{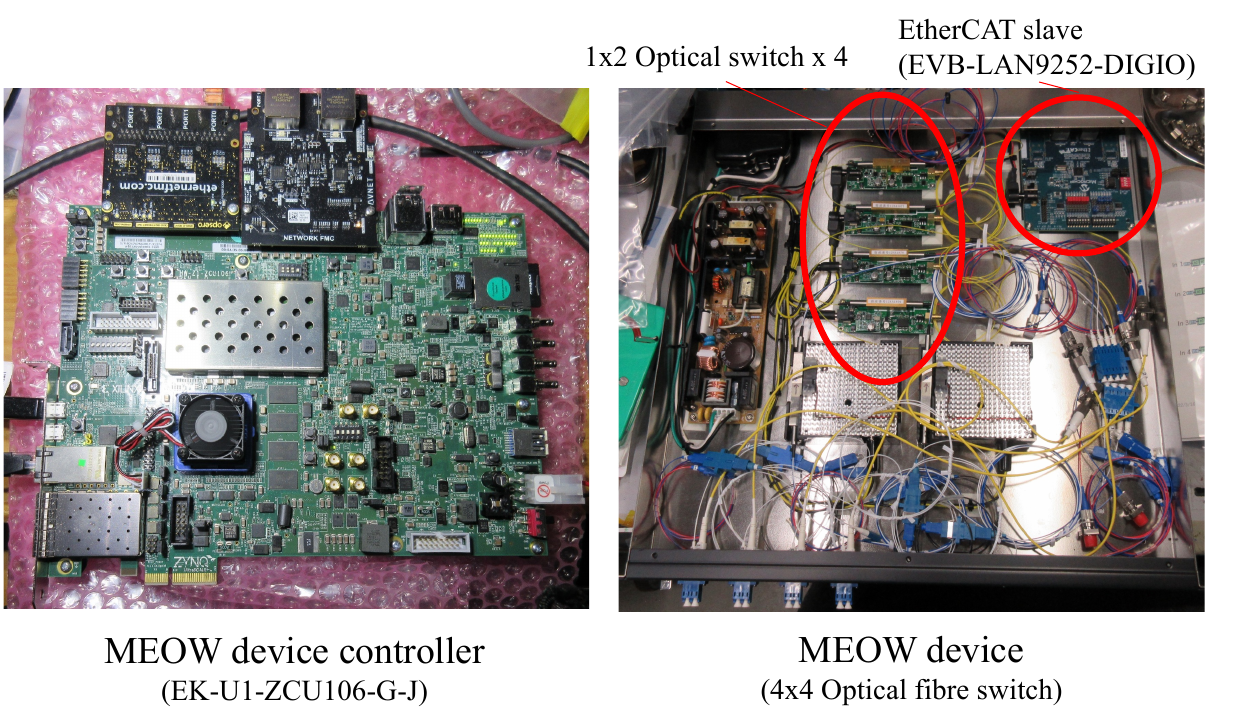}
  \caption{MEOW device controller (left) and MEOW device (right)}
  \label{fig:meow2}
\end{figure}

MEOW consists of three components: network controller, device controller, and device.
MEOW network controller is implemented as an ONOS module. It maintains an optical path 
management table to centrally organize the states of OCSes.
In order to systematically realize configuration management and routing of OCSes, 
we adopt the Functional Block based Disaggregation model~\citet{Ishii2021.JLT}, which describes 
the functions of various optical components in a unified manner.
MEOW device controller manages multiple EtherCAT masters, and MEOW device represents an optical 
switch which has the EtherCAT slave functionality.
Figure~\ref{fig:meow2} shows the reference implementation of MEOW device controller and
MEOW device. The detail will be written in the next section.
Each master controls optical components like wavelength switches and space switches 
through the corresponding slaves.

\subsection{FPGA-based Multi-master EtherCAT system}
Based on our previous experiences~\citet{Takano2021.OCP}, a hardware based EtherCAT
master is critical to meet time constraint and scalability. 
Therefore, we use SoC FPGA to implement the device controller and multi-master
functionalities in a single FPGA chip.

We have implemented the device controller on a Xilinx ZCU106 board,
which provides a Zynq UltraScale+ microprocessor SoC, DDR4 memory, FMC expansion ports,
high-speed serial transceivers, etc.
The SoC consists of the Arm Cortex processor and programmable logic~(PL) in the same chip.
A multi-master EtherCAT functionality is implemented in the PL area, 
and it operates at 100~MHz.
The current implementation supports up to six EtherCAT masters in a single chip.
A real-time Linux is running on the Arm Cortex processor and the intermediate protocol
conversion between the MEOW southbound API and EtherCAT is implemented by combining 
a user-space process and a kernel-space real-time task.

\section{Experiment}

\begin{figure}[t!]
  \centering
  \subfloat[8 devices $\times$ 1 segment\label{fig:expr.setting.1seg}]{%
    \includegraphics[width=.7\linewidth]{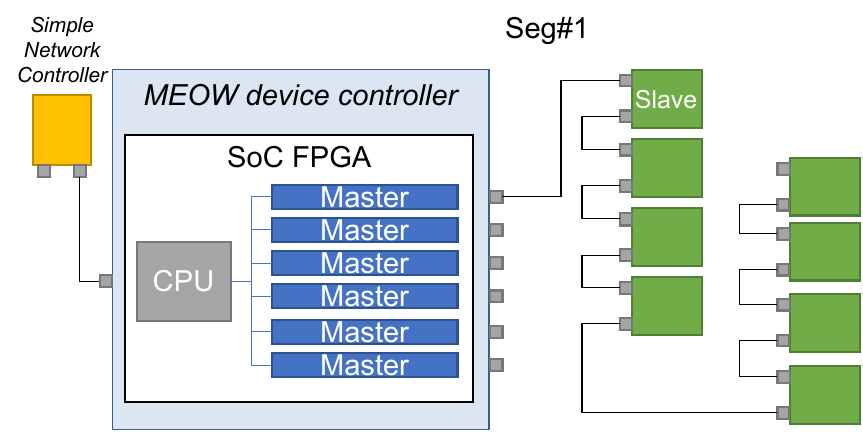}}
  \vspace{4pt}
  \subfloat[2 devices $\times$ 4 segments\label{fig:expr.setting.4seg}]{%
    \includegraphics[width=.7\linewidth]{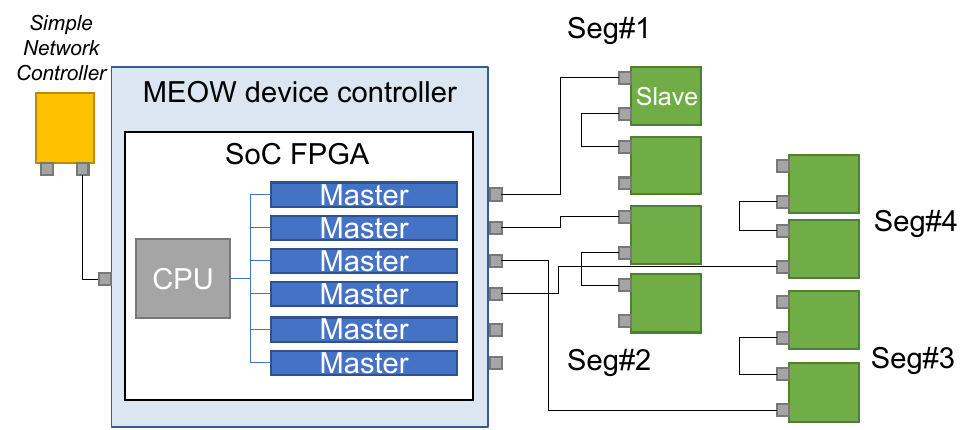}}
  \caption{Experimental Configuration}
  \label{fig:expr.setting}
\end{figure}

In order to demonstrate the feasibility of MEOW, we set up a small experimental
environment equipped with one device controller and eight devices, 
as shown in Figure~\ref{fig:expr.setting}.
We have conducted the experiment in two EtherCAT network topologies. In the first topology 
shown in Figure \ref{fig:expr.setting.1seg}, eight MEOW devices were connected in cascade, 
and we measured the elapsed time from when the controller generated a request to 
when all devices completed the request. In the second topology shown in Figure 
\ref{fig:expr.setting.4seg}, there are four segments and two devices belongs to each segment.

We used Microchip Technology EVB-LAN9252-DIGIO boards as MEOW devices, and they
are connected with the MEOW device controller through 100~Mbps EtherCAT.
This experiment did not use real optical switches. Instead, we have observed a
control signal for configuring an optical switch by using an oscilloscope (Siglent SDS2304X).
Therefore, the assertion of the bit of the most distant device from the controller
was regarded as the completion of the request.
In addition, we implemented a simple network controller on the Raspberry
Pi 4 Model B instead of using an ONOS-based network controller.
This network controller supports the same messaging protocol with a MEOW device controller.

\begin{figure}[t!]
  \centering
  \subfloat[8 devices $\times$ 1 segment. The PDO cycle is 32 {\textmu}s.\label{fig:expr.timing.1seg}]{%
  \includegraphics[width=.7\linewidth]{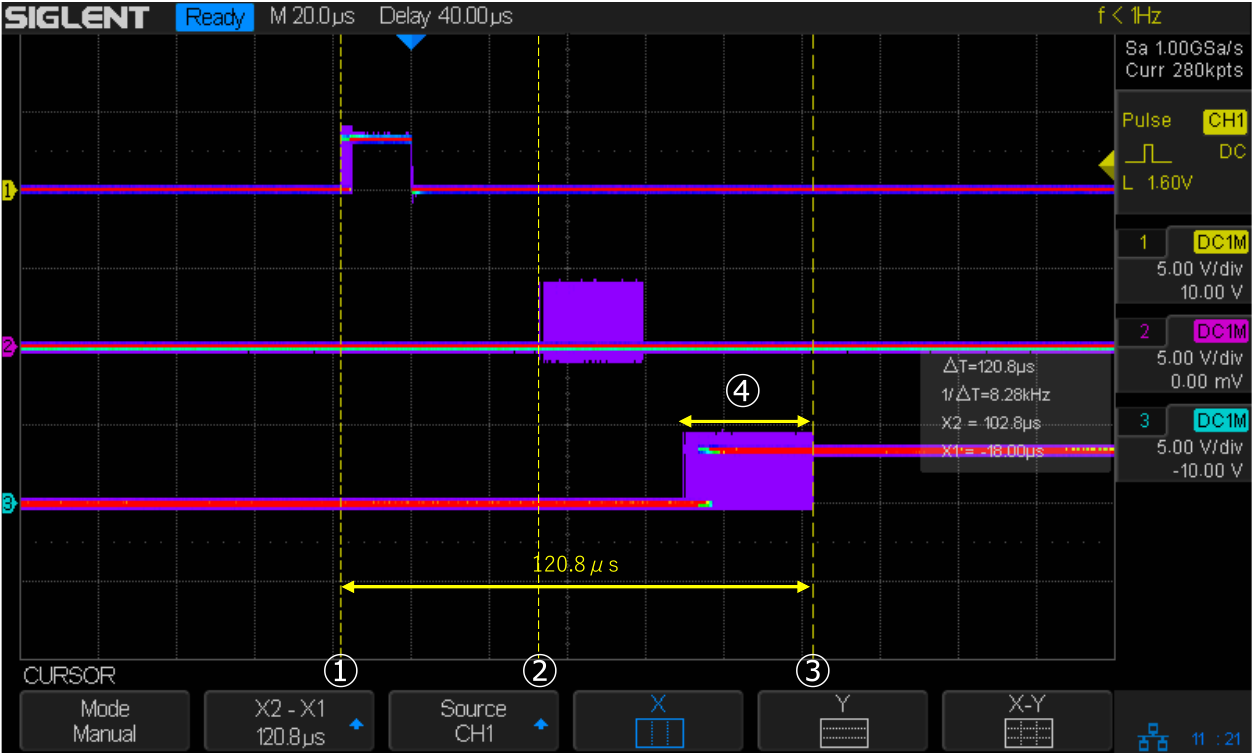}}
  \vspace{8pt}
  \subfloat[2 devices $\times$ 4 segments. The PDO cycle is 80 {\textmu}s.\label{fig:expr.timing.4seg}]{%
  \includegraphics[width=.7\linewidth]{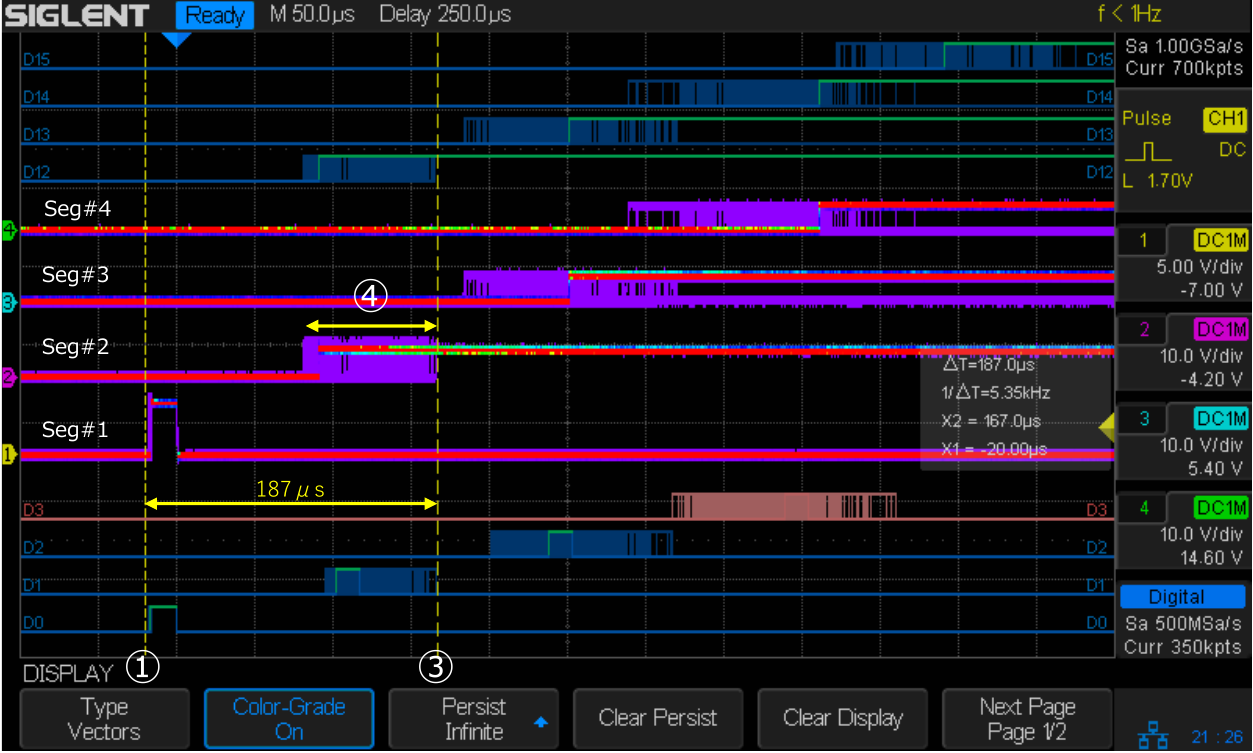}}
  \caption{Sample timing charts. 
    \ctext{1} the simple network controller starts sending the configuration request,
    \ctext{2} the master starts sending the request to slaves,
    \ctext{3} All slaves activate it,
    and \ctext{4} Jitter of the configuration time.}
  \label{fig:expr.timing}
\end{figure}

Figure~\ref{fig:expr.timing} shows the timing chart observed each of 1000 
requests on the above two topologies.
In Figure~\ref{fig:expr.timing.1seg}, the best and the worst 
configuration times of the most distant device from the master are 90 {\textmu}s
and 120.8 {\textmu}s, respectively. 
The configuration time depends on the timing of request reception, and 
in the worst case, it is necessary to wait one Process Data Objects (PDO) cycle
for transmission.
In Figure \ref{fig:expr.timing.4seg}, the best and the worst 
configuration times of 2nd device in the segment \#1 are 100 {\textmu}s 
and 187 {\textmu}s, respectively.

The experimental result in Figure \ref{fig:expr.setting.1seg} shows that the 
configuration time increases by 0.9 {\textmu}s for each more device in the single segment.
When controlling 1000 racks divided into four segments, the worst configuration time
can be 412 {\textmu}s by extrapolating from the experimental results.
The current MEOW device controller has implemented six EtherCAT masters. When we utilize 
six masters, the worst configuration time can decrease to 350 {\textmu}s.

There is room to further shorten the configuration time. A faster data transmission 
(e.g., 1 Gbps) is promising. Increasing parallelism, in other words, increasing the 
number of device controllers and decreasing the number of devices in a segment,
also helps for the purpose.
Moreover, we will investigate how to achieve stable multi-master operation with 
the PDO cycle as short as that of single-master operation. If the PDO cycle can
be reduced as in the single-master operation, the worst configuration time can
be reduced by 50 {\textmu}s.

\section{Conclusion}

This paper proposes MEOW, a fast control plane technology for large-scale and high-speed 
OCSes. MEOW is leveraged by the combination of commodity hardware and software,
including an industrial real-time Ethernet standard EtherCAT, FPGA implementation of 
multiple EtherCAT masters, and an open source SDN controller ONOS.
The proposed technology is expected to pave the way for future application of high-speed
OCSes in data center networks.

The source code of MEOW network controller is available from \url{https://github.com/ryousei/meow-rest/}.

\section*{Acknowledgement}
This paper is based on results obtained from a project, JPNP16007, commissioned
by the New Energy and Industrial Technology Development Organization (NEDO).

\bibliographystyle{unsrtnat}
\bibliography{meow}

\begin{thebibliography}{6}
\providecommand{\natexlab}[1]{#1}
\providecommand{\url}[1]{\texttt{#1}}
\expandafter\ifx\csname urlstyle\endcsname\relax
  \providecommand{\doi}[1]{doi: #1}\else
  \providecommand{\doi}{doi: \begingroup \urlstyle{rm}\Url}\fi

\bibitem[Takano et~al.(2021)Takano, Shimizu, Okazaki, Ishii, Namiki, and Sato]{Takano2021.OCP}
Ryousei Takano, Toshiyuki Shimizu, Fumihiro Okazaki, Kiyo Ishii, Shu Namiki, and Kenichi Sato.
\newblock {Demonstration of Multimaster EtherCAT-based control plane for Optical netWorks (MEOW)}.
\newblock In \emph{2021 Open Compute Project Future Technologies Symposium}, pages 1--2. OCP, November 2021.

\bibitem[Farrington et~al.(2011)Farrington, Porter, Radhakrishnan, Bazzaz, Subramanya, Fainman, Papen, and Vahdat]{Farrington2011.SIGCOM}
Nathan Farrington, George Porter, Sivasankar Radhakrishnan, Hamid~Hajabdolali Bazzaz, Vikram Subramanya, Yeshaiahu Fainman, George Papen, and Amin Vahdat.
\newblock Helios: a hybrid electrical/optical switch architecture for modular data centers.
\newblock \emph{ACM SIGCOMM Computer Communication Review}, 41\penalty0 (4):\penalty0 339--350, 2011.

\bibitem[Poutievski et~al.(2022)Poutievski, Mashayekhi, Ong, Singh, Tariq, Wang, Zhang, Beauregard, Conner, Gribble, Kapoor, Kratzer, Li, Liu, Nagaraj, Ornstein, Sawhney, Urata, Vicisano, Yasumura, Zhang, Zhou, and Vahdat]{Poutievski2022.SIGCOMM}
Leon Poutievski, Omid Mashayekhi, Joon Ong, Arjun Singh, Mukarram Tariq, Rui Wang, Jianan Zhang, Virginia Beauregard, Patrick Conner, Steve Gribble, Rishi Kapoor, Stephen Kratzer, Nanfang Li, Hong Liu, Karthik Nagaraj, Jason Ornstein, Samir Sawhney, Ryohei Urata, Lorenzo Vicisano, Kevin Yasumura, Shidong Zhang, Junlan Zhou, and Amin Vahdat.
\newblock Jupiter evolving: Transforming google's datacenter network via optical circuit switches and software-defined networking.
\newblock In \emph{Proceedings of the ACM SIGCOMM 2022 Conference}, page 66^^e2^^80^^9385. ACM, 2022.
\newblock ISBN 9781450394208.
\newblock \doi{10.1145/3544216.3544265}.
\newblock URL \url{https://doi.org/10.1145/3544216.3544265}.

\bibitem[{K. Sato}(2018)]{Sato2019.JLT}
{K. Sato}.
\newblock Realization and application of large-scale fast optical circuit switch for data center networking.
\newblock \emph{Jornal of Lightwave Technology}, 36\penalty0 (7):\penalty0 1411--1419, April 2018.

\bibitem[Berde et~al.(2014)Berde, Gerola, Hart, Higuchi, Kobayashi, Koide, Lantz, O'Connor, Radoslavov, Snow, et~al.]{Berde2014.HotSDN}
Pankaj Berde, Matteo Gerola, Jonathan Hart, Yuta Higuchi, Masayoshi Kobayashi, Toshio Koide, Bob Lantz, Brian O'Connor, Pavlin Radoslavov, William Snow, et~al.
\newblock {ONOS: towards an open, distributed SDN OS}.
\newblock In \emph{Proceedings of the third workshop on Hot topics in software defined networking}, pages 1--6. ACM, 2014.

\bibitem[Ishii et~al.(2021)Ishii, Xu, Yoshikane, Takefusa, Tsuritani, Awaji, and Namiki]{Ishii2021.JLT}
Kiyo Ishii, Sugang Xu, Noboru Yoshikane, Atsuko Takefusa, Takehiro Tsuritani, Yoshinari Awaji, and Shu Namiki.
\newblock Automatic mapping between real hardware composition and roadm model for agile node updates.
\newblock \emph{Journal of Lightwave Technology}, 39\penalty0 (3):\penalty0 821--832, 2021.
\newblock \doi{10.1109/JLT.2020.3048424}.

\end{thebibliography}

\end{document}